\documentclass[twocolumn,superscriptaddress,showpacs,aps,prl]{revtex4}
\usepackage{makeidx}
\usepackage{bm}
\usepackage{graphics}
\usepackage[dvips]{epsfig}
\newcounter{fig}

\begin{document}
\title{ Gate-tunable bandgap in bilayer graphene}
\author{L.A. Falkovsky}
\affiliation{L.D. Landau Institute for Theoretical Physics, Moscow
117334, Russia} \affiliation{Institute of the High Pressure
Physics, Troitsk 142190, Russia}

\date{\today}      

\begin{abstract}
The tight-binding model of  bilayer graphene  is used to find the
gap between the conduction and valence bands, as a function of
both the gate voltage and as  the doping by donors or acceptors.
The total Hartree energy is minimized and the equation for the gap
is obtained. This equation for the ratio of the gap to the
chemical potential is  determined  only by the screening constant.
Thus the gap is strictly proportional to the gate voltage or the
carrier concentration in  the absence of donors or acceptors. In
the opposite  case, where the donors or acceptors are present, the
gap demonstrates the asymmetrical behavior on the electron and
hole sides of the gate bias. A comparison with  experimental data
obtained by Kuzmenko et al demonstrates the good agreement.
\end{abstract}
\maketitle
\section{Introduction}
Bilayer graphene has attracted much interest partly due to the
opening of a tunable gap in its electronic spectrum by  an
external electrostatic field. Such a phenomenon was predicted in
Refs.  \cite{McF,LCH} and can be observed in optical studies
controlled  by applying a gate bias
\cite{OBS,ZBF,KHM,LHJ,ECNM,NC,MLS,KCM}. In  Refs. \cite{Mc,MAF},
within the self-consistent Hartree approximation, the gap was
derived  as a near-linear function of the carrier concentration
injected in the bilayer by the gate bias.
Recently, this problem was
 numerically considered in Ref. \cite{GLS} using the density functional theory (DFT)
 and including the external charge doping involved with impurities.
 The DFT calculation
 gives the gap  which is roughly  half of the  gap
 obtained in the Hartree approximation. This disagreement was
 explained in Ref. \cite{GLS} as a result of both the inter- and intralayer
 correlations.

 In this work, we study this problem within the same
 Hartree approximation as in Refs. \cite{Mc,MAF}, but including
 the effect of  external doping. We calculate  the  carrier
 concentration on  both sides of the bilayer
 considering  the case, where the carrier
 concentration in the bilayer is  less than 10$^{13}$ cm$^{-2}$.  Then,
  we minimize the total energy of the system and find self-consistently
  both the chemical potential and the gap induced by the gate bias.
  Our results completely differ from those
 obtained  in Refs. \cite{Mc,MAF}, where  the external
 doping is disregarded. In the presence of dopants, the dependence of
 the gap on the carrier concentration, i.\,e., on the gate voltage,
 exhibits an asymmetry at the electron and hole sides of the gate bias.
\section{Tight-binding model of bilayer graphene}
\begin{figure}[h]
\resizebox{.25\textwidth}{!}{\includegraphics{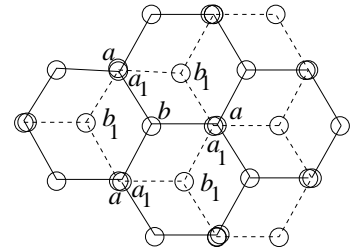}}
\caption{Bilayer lattice} \label{grlat}
\end{figure}

 The  graphene bilayer lattice is shown in Fig. \ref{grlat}.
Atoms  in one layer, i.\,e., $a$ and $b$ in the unit cell, are
connected by  solid lines, and in the other layer, e.\,g., $a_1$
and $b_1$, by the dashed lines. The atom $a$ ($a_1$) differs from
$b$ ($b_1$) because it has a neighbor just below in the adjacent
layer, whereas the atom $b$ ($b_1$) does not.

Let us recall the main results of the Slonchewski--Weiss--McClure
model \cite{SW,McCl}. In the tight-binding model, the Bloch
functions of the bilayer are written as
\begin{eqnarray}\nonumber \label{eh}
\psi_a=\frac{1}{\sqrt{N}}\sum_{j}e^{i{\bf ka}_j}\psi_0({\bf
a}_j-{\bf r})\\ \psi_b=\frac{1}{\sqrt{N}}\sum_{j}e^{i{\bf
ka}_j}\psi_0({\bf a}_j+{\bf a}-{\bf r})\\ \nonumber
\psi_{a1}=\frac{1}{\sqrt{N}}\sum_{j}e^{i{\bf ka}_j}\psi_0({\bf a}_j+{\bf c}-{\bf r})\\
\psi_{b1}=\frac{1}{\sqrt{N}}\sum_{j}e^{i{\bf ka}_j}\psi_0({\bf
a}_j+{\bf c}+{\bf a}-{\bf r}),\nonumber
\end{eqnarray}
where the sums are taken over the  lattice vectors ${\bf a}_j$ and
$N$ is the number of unit cells. Vectors  ${\bf a}$ and ${\bf c}$
connect the nearest atoms in the layer and in the neighbor layers,
correspondingly.

 For the nearest neighbors, the effective Hamiltonian in the space
 of the functions (\ref{eh}) can be written as
 \begin{equation}
H(\mathbf{k})=\left(
\begin{array}{cccc}
U+\Delta \,    & \gamma_0f^{*} \,& \gamma_1    \, & \gamma_4f\\
\gamma_0f \,& U-\Delta     \, & \gamma_4f\,& \gamma_3f^{*}\\
\gamma_1    \,  &\gamma_4f^{*} \,&-U+\Delta  \,  &\gamma_0f\\
\gamma_4f^{*} \,&  \gamma_3f\,&\gamma_0f^{*} \,&-U-\Delta
\end{array}%
\right) ,  \label{haml}
\end{equation}%
where $f=\gamma_0 \left[ e^{ik_{x}a}+2e^{-ik_{x}a/2}\cos {(k_{y}a%
\sqrt{3}/2)}\right] $.  The values of hopping integrals
$\gamma_0,\gamma_1,\gamma_3,\gamma_4,$ and $\Delta$ are given in
the Table I.
\begin{table}[h]
\caption{\label{tb1} Parameter values of the electron spectrum in
eV. }
        \begin{ruledtabular}
                \begin{tabular}{|c|c|c|}
Parameter     &Experiment \cite{KCM} &DFT calculation \cite{CM}\\
\hline$\gamma_0$&$3.16\pm0.3$& $2.598\pm0.015$ \\
\hline$\gamma_1$&$0.381\pm0.003$& $0.34\pm0.02$ \\
\hline$\gamma_3$&$0.38\pm0.06$& $0.32\pm0.02$ \\
\hline$\gamma_4$&$0.14\pm0.03$& $0.177\pm0.025$ \\
\hline$\Delta$&$0.022\pm0.003$& $0.024\pm0.01$ \\
\end{tabular}
\end{ruledtabular}
\end{table}
The largest of them, $\gamma_0$, determines the band dispersion
near the $K$ point in the Brillouin zone where the matrix element
$\gamma_0f$ can be expanded as
\[
\gamma_0f=v(ik_{x}-k_{y}),
\]%
with a velocity parameter $v=3\gamma_0 a/2$.
 The parameters $\gamma_3$ and $\gamma_4$ giving a correction
to the dispersion are  less than $\gamma_0$ by a factor of 10. The
parameters $\gamma_1$ and $\Delta$ result in the position  of
levels at $K$, but $\Delta$ is much less than $\gamma_1$.  There
is in addition the parameter $U$ induced by the gate voltage and
associated with the asymmetry  of two layers in the external
electrostatic field. This parameter  presents the potential energy
$-edE$ between two layers, where $d$ is the interlayer distance
and $E$ is electric field induced both by the gate voltage and the
external  dopants in the bilayer.

The parameter $U$ as well as the chemical potential $\mu$ should
be self-consistently calculated  for the given gate voltage. For
this purpose, we can held only the parameters $\gamma_0$ and
$\gamma_1$, neglecting the small effect of $\gamma_3$, $\gamma_4$,
and $\Delta$ on the gap $U$.  In this approximation, the effective
Hamiltonian can be written in the simple form
\begin{equation}
H(\mathbf{k})=\left(
\begin{array}{cccc}
U \,    & vk_{+} \,& \gamma_1    \, & 0\\
vk_{-} \,& U     \, & 0\,& 0\\
\gamma_1    \,  &0 \,&-U  \,  &vk_{-}\\
0 \,& 0 \,&vk_{+} \,&-U
\end{array}%
\right) ,  \label{ham}
\end{equation}%
where $k_{\pm}=\mp ik_x-k_y$ in the vicinity of the $K$ points.

\begin{figure}[h]
\resizebox{.4\textwidth}{!}{\includegraphics{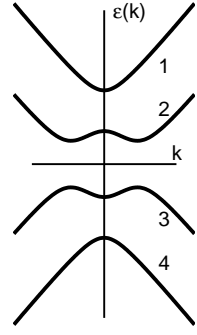}}
\caption{Band structure of bilayer.} \label{dis}
\end{figure}

 The Hamiltonian gives four energy bands:
\begin{eqnarray}\label{bands}
\varepsilon_{1,4}(q)=\pm\left(\frac{\gamma_1^2}{2}+U^2+q^2+W\right)^{1/2}\,,\\\nonumber
\varepsilon_{2,3}(q)=\pm\left(\frac{\gamma_1^2}{2}+U^2+q^2-W\right)^{1/2}\,,
\end{eqnarray}
where
$$W=\left(\frac{\gamma_1^4}{4}+(\gamma_1^2+4U^2)q^2\right)^{1/2}$$
and we denote  $q^2=(vk)^2$.

 The band structure is shown in Fig.\,
\ref{dis}. The minimal value of the upper  energy $\varepsilon_1$
is $\sqrt{U^2+\gamma_1^2}$. The
 $\varepsilon_2$ band takes the maximal value $|U|$ at $q=0$ and
the minimal value $\tilde{U}=\gamma_1|U|/\sqrt{\gamma_1^2+4U^2}$
at $q^2=2U^2(\gamma_1^2+2U^2)/(\gamma_1^2+4U^2).$ Because the
value of $U$ is much less than $\gamma_1$, the distinction between
$U$ and $\tilde{U}$ is small and the gap between the bands
$\varepsilon_2$ and $\varepsilon_{3}$ takes approximately the
value $2|U|$.

\section{Eigenfunctions and carrier concentration }

 The four eigenfunctions ${\mathbf C}$ corresponding with eigenvalues
 (\ref{bands}) of
the Hamiltonian (\ref{ham}) read
\begin{equation}
{\mathbf C}=\frac{1}{C}\left(
\begin{array}{c}
(U-\varepsilon_n)[(\varepsilon_n+U)^2-q^2]\\
-q_{-}[(\varepsilon_n+U)^2-q^2]\\
\gamma_1(U^2-\varepsilon_n^2)\\
\gamma_1q_{+}(U-\varepsilon_n)\end{array}\right) ,\label{fun}
\end{equation}
where the ${\mathbf C}$ norm squared is
\begin{eqnarray}\nonumber
C^2=[(\varepsilon_n+U)^2-q^2]^2[(\varepsilon_n-U)^2+q^2]\\\nonumber
+\gamma_1^2(\varepsilon_n-U)^2
[(\varepsilon_n+U)^2+q^2]\,.\end{eqnarray} As seen from Eqs.
(\ref{eh}), the probability $p_1$ to find an electron, for
instance, on the first layer is \[p_1=|C_1|^2+|C_2|^2\,,\]  where
the subscript $C_i$ numerates the elements of the column
(\ref{fun}).

We assume, that  carriers  occupy only the bands
$\varepsilon_{2,3}$, so the chemical potential $\mu$ and the gap
$2|U|$ are  less than the distance between the bands
$\varepsilon_1$ and $\varepsilon_2$, i.\,e., $(|\mu|,
2|U|)<\gamma_1.$ The electron dispersion for  the
$\varepsilon_{2,3}$ bands  can be expanded in powers of $q^2$:
\begin{equation}\nonumber
\varepsilon_{n}^2(q)=U^2-4\frac{U^2}{\gamma_1^2}q^2+\frac{q^4}
{\gamma_1^2}\,,
\end{equation}
where $n=2$  stands for the electron conductivity and $n=3$ for
the hole conductivity. Then, for $q^2\gg 4U^2$, we can omit here
the second term and use the simple relations
\begin{eqnarray}\label{q}
q^2=\gamma_1(\varepsilon_n^2-U^2)^{1/2}
\end{eqnarray}
  neglecting the small effect of the "mexican hat".

 Keeping only
the leading terms, one can find with the help of Eq. (\ref{fun}),
that the probabilities $p_{1,2}$ to find an electron on the layers
are proportional
\begin{eqnarray}\nonumber
p_1=|C_1|^2+|C_2|^2\propto q^6
=\gamma_1^3(\varepsilon_{n}^2-U^2)^{3/2}\,,\\
\nonumber p_2= |C_3|^2+|C_4|^2\propto
q^2\gamma_1^2(U-\varepsilon_{n})^{2}\\ \nonumber
=\gamma_1^3(\varepsilon_{n}^2-U^2)^{1/2}(U-\varepsilon_{n})^2\,.
\end{eqnarray}
Therefore, the normalized probability to find an electron, for
instance, on the first layer  can be written as
\begin{eqnarray}\nonumber
p_1=\frac{(\varepsilon_{n}^2-U^2)^{3/2}}
{(\varepsilon_{n}^2-U^2)^{3/2}+
(\varepsilon_{n}^2-U^2)^{1/2}(U-\varepsilon_{n})}\\ \label{prob1}
=(\varepsilon_{n}+U)/2\varepsilon_{n} \,.
\end{eqnarray}
Within  the approximation (\ref{q})-(\ref{prob1}), many observable
effects can be analytically evaluated for the intermediate carrier
concentration, $4U^2\ll\gamma_1\sqrt{\mu^2-U^2}\ll\gamma_1^2$.

At zero temperature, the carrier concentration on the sides of the
bilayer is found with the help of Eq. (\ref{prob1}) as
\begin{eqnarray}\nonumber
n_{1,2}=\frac{2}{\pi\hbar^2 v^2}\int p_{1,2}\, qdq\\
\label{n1}
=\frac{n_0U}{2\gamma_1}[\sqrt{x^2-1}\pm \ln{(x+\sqrt{x^2-1}}]\,,
\end{eqnarray}
where the limits of integration are $q=0$ and the chemical
potential $\mu$ and we set
\begin{eqnarray}\nonumber
n_0=\gamma_1^2/\pi\hbar^2v^2=1.03\times 10^{13} \text{cm}^{-2}\,,\\
 x=|\mu/U|.
\label{n0} \end{eqnarray} For the total carrier concentration $n$
in the bilayer,
we obtain
\begin{eqnarray}\label{n}
n=\frac{\gamma_1}{\pi\hbar^2v^2}\sqrt{\mu^2-U^2}
=\frac{n_0U}{\gamma_1}\sqrt{x^2-1}\,.
\end{eqnarray}
\section{Minimization of the total energy}
\begin{figure}[h]
\resizebox{.2\textwidth}{!}{\includegraphics{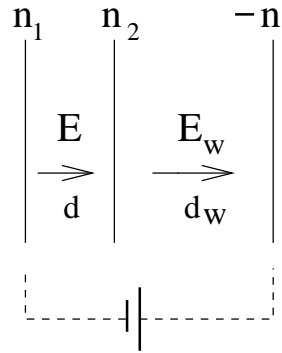}}
\caption{Electrostatic model; $d$ is the interlayer distance,
$d_w$ is the wafer thickness.} \label{bifig1}
\end{figure}
In order to find the chemical potential $\mu$ and the gap $2|U|$
at the given gate voltage
\begin{equation}\label{con}
eV_g=-edE-ed_wE_w\,,
\end{equation}
 we minimize the total energy containing
both the energy $V^{(c)}$ of the carriers and the energy $V^{(f)}$
of the electrostatic field. Within the Hartree approximation, when
no electron  correlations are taken into account, the filled bands
do not contribute into the  energy of the system, because the
electron charge of the filled bands is compensated by the ion
charge and this energy have to be considered as the ground state
energy. The excitation  energy is owes its origin to the carriers
in unfilled bands.  Electrons in the $\varepsilon_2$ band or holes
in the $\varepsilon_3$ band contribute in the total energy of the
system the energy
 \begin{eqnarray} \nonumber\label{wk}
V^{(c)}=\frac{2}{\pi\hbar^2 v^2}\int |\varepsilon_{n}(q)|qdq\\
=\frac{n_0 U^2}{2\gamma_1}
[x\sqrt{x^2-1}+\ln{(x+\sqrt{x^2-1})}]\,.
\end{eqnarray}
 The energy of the electrostatic field  (see Fig. \ref{bifig1})
\begin{equation}\label{wf}
V^{(f)}=\frac{1}{8\pi}(dE^2+\epsilon_w d_w E^2_w)
\end{equation}
 can be  written in
terms of the carrier concentrations with the help of relations
\begin{equation}\label{ne}
4\pi e(n_1-N_1)=E\quad  \text{and}\quad 4\pi e(n-N)=\epsilon_w
E_w\,,
\end{equation}
where $\epsilon_w$ is the dielectric constant of the wafer, $N_1$
and $N_2$ are  concentrations of the acceptor or donor impurities
on the left and right layers, correspondingly, whereas  the total
dopant concentration on the bilayer is $N=N_1+N_2$.  All these
numbers are supposed to be positive or negative for the electron
or hole doping correspondingly. Let us emphasize that the
dielectric constant $\epsilon$ of bilayer graphene depends on the
substrate. For simplicity, we put $\epsilon=1$ in the definition
(\ref{wf}).

 We seek the minimum of the total energy
$$V^{(f)}+V^{(c)}+\lambda(eV_g+edE+ed_wE_w)$$
as a function of $U$, $\mu$, and the Lagrange multiplier
$\lambda$. Differentiation with respect to $\lambda$ gives the
gate voltage constraint (\ref{con}). Minimization  with respect to
$U$ and $x$ gives
\begin{eqnarray} \nonumber
4\pi e^2[n_1-N_1)n_{1u}d+(n-N)n_ud_w/\epsilon_w\\
\nonumber +V^{(c)}_u+ 4\pi e^2\lambda(n_{1u}d+n_ud_w/\epsilon_w)=0
\end{eqnarray}
and   the similar  equation with a substitution $u\rightarrow x$,
where the subscripts $u$ and $x$ note the derivatives with respect
to the corresponding variables. The Lagrange multiplier $\lambda$
can be excluded from these two equations. Then, the equation
obtained should be expanded in $d/d_w$, since the thickness $d$ of
the bilayer is much less than the thickness $d_w$ of the
substrate.

Thus, we obtain the following equation:
\begin{equation}\label{var}
4\pi
e^2d\left(n_2-N_2\right)\left(\frac{n_{1x}}{n_x}-\frac{n_{1u}}{n_u}\right)=
\frac{V^{(c)}_x}{n_x}-\frac{V^{(c)}_u}{n_u}\,.
\end{equation}
Let us emphasize, that this equation  is invariant  under the
simultaneous sign change  in $n_{1,2}$ and $N_{1,2}$, that
expresses the charge invariance of the problem. At the fixed
external doping $N_{1,2}$, the gap on the electron and hole sides
of the gate bias is not symmetrical.

The derivatives in Eq. (\ref{var}) are calculated with the help of
Eqs. (\ref{n1})--(\ref{ne}). As a result, Eq. (\ref{var}) becomes
\begin{eqnarray}\label{var1}
2\frac{\gamma_1 N_2}{Un_0}=\sqrt{x^2-1}\pm\left\{f(x)+\frac{x
f(x)}{\Lambda[x f(x)-\sqrt{x^2-1}]}\right\}
\end{eqnarray}
with the function $f(x)=\ln{(x+\sqrt{x^2-1})}$  and the
dimensionless screening constant
\begin{equation}\label{lam}
\Lambda=\frac{e^2\gamma_1d}{(\hbar v)^2}\,.
\end{equation}
  For the parameters of graphene $d=3.35\, \AA\,, \gamma_1=0.381$ eV, and
$v=1.02\times 10^8$ cm/s, we get $\Lambda=0.41$.

\section{The gap in undoped and doped based bilayer}
\subsection{(i) undoped bilayer}
\begin{figure}[]
\resizebox{.5\textwidth}{!}{\includegraphics{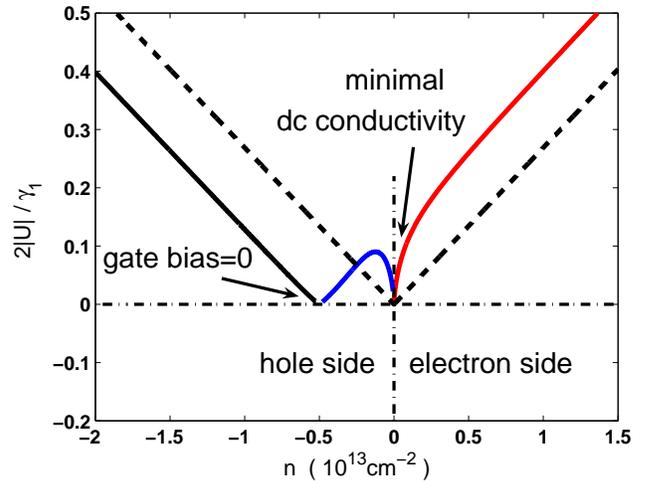}}
\caption{The gap in units of $\gamma_1=0.381 $ eV versus the
carrier concentration in  absence of doping (dashed line) and for
the hole doping level $N_2=-2.5\times10^{12}$ cm$^{-2}$ (solid
line); the positive (negative) values of $n$ correspond to the
electron (hole) conductivity. The difference between values of $n$
marked as "gate bias = 0" and "minimal dc conductivity" is
$2N_2$.} \label{bifig3}
\end{figure}

First, let us consider  an ideal undoped  bilayer  with
$N_1=N_{2}=0$.
We get a nonzero  solution for $U$, if the right-hand side of Eq.
(\ref{var1}) vanishes. This condition  is fulfilled only for the
sign "--" in Eq. (\ref{var1}), that defines the polarity of the
layers [see Eq. (\ref{n1})]. We obtain solution $x=x_0=6.61$.
 According to Eq. (\ref{n}), the gap as a function of the carrier
 concentration takes a very simple form:
\begin{equation} \label{w}
 2|U/n|=\frac{2\gamma_1}{n_0\sqrt{x_0^2-1}}=1.13\times10^{-11}
\text{meV} \cdot \text{cm}^2\,,
\end{equation}
where the right-hand side does not depend  at all on the gate
bias, but only on the screening constant $\Lambda$. This
dependence is shown in Fig. \ref{bifig3} in  dashed lines, it is
symmetrical on the electron and hole sides.

With the help of  Eq. (\ref{n0}), we obtain the chemical potential
as a linear function of the carrier concentration
\begin{equation} \label{mu}
 \mu=\frac{\gamma_1x_0}{n_0\sqrt{x_0^2-1}}n\,,
\end{equation}
where $n$ is positive (negative) for the electron (hole)
conductivity.

We can compare Eq. (\ref{w}) with the corresponding result of Ref.
\cite{Mc}:
\begin{equation}\label{w1}
2|U/n|=\frac{e^2d}{2\epsilon_0}\left[1+2\Lambda\frac{|n|}{n_0}+
\Lambda\ln\frac{n_0}{|n|}\right]^{-1}\,.
\end{equation}
Both equations give approximately the same results   at $|n|\simeq
0.1 n_0\simeq 10^{12}$ cm$^{-2}$. However, contrary to Eq.
(\ref{w}), Eq. (\ref{w1}) contains the carrier concentration in
the right-hand side giving rise to the more rapid increase in the
gap with $|n|\ll n_0$. This increase also contradicts to the DFT
calculations \cite{GLS}.

Two reasons can result  in the disagreement of our theory with
Ref. \cite{Mc}. First, in Ref. \cite{Mc}, the filled bands are
supposed to create the electric field in  the bilayer, that is
incorrect as explained in the previous section. Second, the
minimization should be done with respect two variables $\mu$ and
$U$, however, only one of them seems to be exploited in Ref.
\cite{Mc}.

\subsection{(ii) doped bilayer}
 For the bilayer with the acceptor or donor dopants,
  Eq. (\ref{var1}) presents a solution $w=2\gamma_1N_2/Un_0$ as a
function of $x$. We obtain, evidently, the small values of $w$ for
$x$ close to $x_0=6.61$. Since $x_0\gg 1$, we can expand the
function in the right-hand of Eq. (\ref{var1}) in $1/x$. In this
region of the relatively large $|U|$, we find again with the help
of Eqs. (\ref{n}) and (\ref{var1}) the linear dependence
\begin{eqnarray}\label{nd}
 2|U|=|n-2N_2|\frac{2\gamma_1}{n_0x_0}\\
\nonumber
=1.13\,|n-2N_2| \times10^{-11} \text{ meV}\cdot
\text{cm}^2.
\end{eqnarray}

 The value of the carrier concentration $n=2N_2$ corresponds to
the zero bias voltage, where $U=0$ (see Fig. \ref{bifig3}).
Therefore, in contrast with the undoped case, the gap demonstrates
the asymmetrical behavior  on the electron and hole sides.
  If the bilayer contains acceptors with concentration $N_2$, the gap decreases
linearly with the hole concentration and vanishes, when the gate
bias is not applied and the hole concentration equals $2N_2$.
Starting from this point, the gap
  increases and, thereafter, becomes  again small
  (equals  zero in Fig. \ref{bifig3}) at the carrier concentration
  corresponding to the minimal value of the dc conductivity, where $n=0$.
Therefore, the  difference ($1.56\times10^{12}$cm$^{-2}$ in Fig.
\ref{bifig3}) observed in Refs. \cite{MLS} and \cite{KCM} between
these two values of carrier concentrations, at the zero bias and
at the minimal conductivity, gives directly the donor/acceptor
concentration ($2N_2$) on the layer close to the  substrate. Then,
for the gate bias applied in order to increase the electron
concentration, the gap is rapidly opening with the electron
appearance.

\begin{figure}[h]
\resizebox{.5\textwidth}{!}{\includegraphics{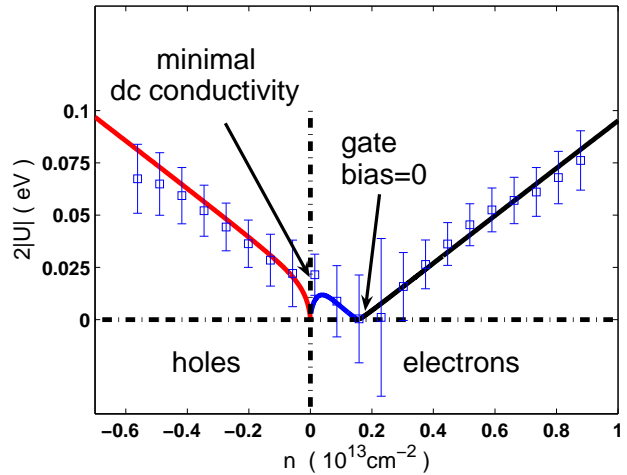}}
\caption{The gap in  eV versus the carrier concentration for the
electron doping with the concentration $N_2=0.78\times10^{12}$
cm$^{-2}$ (our theory); the positive (negative) values of $n$
correspond to the electron (hole) conductivity; squares are
experimental data \cite{KCM}.} \label{bifig4}
\end{figure}

 We see, that the asymmetry arises
between the electron and hole sides of the gate bias. This
asymmetry can simulate a result of  the hopping integral $\Delta$
in the electron spectrum \cite{CNM}. In order to obtain the gap
dependence for the case of electron doping, $N_2>0$, the
reflection transformation $n\rightarrow -n$ has to be made. This
case is shown in  Fig. \ref{bifig4} where the experimental data
from Ref. \cite{KCM} are displayed.

The gap in the vicinity of the minimal conductivity value could
reach indeed a finite value due to several reasons. One of them is
the form of the "mexican hat" shown in Fig. \ref{dis}. Second, the
trigonal warping is substantial at low carrier concentrations.
Finally, the graphene electron spectrum is unstable with respect
to the Coulomb interaction at the low momentum values. For the
graphene monolayer, as shown in Ref. \cite{Mi}, the logarithmic
corrections appear at the small momentum. In the case of the
bilayer, the electron self-energy contains the linear corrections,
as can be found using the perturbation theory. The similar linear
terms resulting in a nematic order were also obtained  in the
framework of the renormalization group \cite{VY}.

\section{Conclusion}
The gap  opening in the gated graphene bilayer has an intriguing
behavior as a function of carrier concentration. In the presence
of the external doping charge, i.\,e. donors or acceptors, this
function is asymmetric on the hole and electron sides of the gate
bias and it is the linear function only for the large gate bias. A
difference between  two values of carrier concentrations, i.\,e.
at the zero bias and  at the minimal conductivity, gives directly
the sign and concentration of the charged dopants on the bilayer.

I thank  A.M. Duygaev and Y.N. Ovchinnikov for helpful discussions
and A.B. Kuzmenko for providing his experimental data
prior to publication. This work was supported by the Russian
Foundation for Basic Research (grant No. 07-02-00571). The author
is grateful to the Max Planck Institute for the Physics of Complex
Systems for hospitality in Dresden.

\end{document}